\documentclass{statsoc}

\usepackage{amsmath,amssymb}
\usepackage{setspace}
\usepackage{graphicx}
\usepackage{hyperref}
\usepackage{bm}
\usepackage{multirow}

\usepackage[paperwidth=8.5in,paperheight=11in,top=1in, bottom=1in, left=1in, right=1in]{geometry}
\usepackage{color}

\newcommand{\Var}{\text{var}}
\newcommand{\var}[1]{\Var \! \left \{ #1 \right \} }

\newcounter{thm}
\def\claim#1{\medskip\par\noindent\hangindent=\parindent \hangafter=0\refstepcounter{thm}{\it Theorem \arabic{thm}. }
}
\def\endclaim{
\par\medskip}
\newenvironment{thm}{\claim}{\endclaim}

\newcounter{cor}
\def\corclaim#1{\medskip\par\noindent\hangindent=\parindent \hangafter=0\refstepcounter{cor}{\it Corollary \arabic{cor}. }
}
\def\endcorclaim{
\par\medskip}

\newcounter{prop}
\def\propclaim#1{\medskip\par\noindent\hangindent=\parindent \hangafter=0\refstepcounter{prop}{\it Proposition \arabic{prop}. }
}
\def\endpropclaim{
\par\medskip}
\newenvironment{prop}{\propclaim}{\endpropclaim}

\newcounter{lemma}
\def\lemmaclaim#1{\medskip\par\noindent\hangindent=\parindent \hangafter=0\refstepcounter{lemma}{\it Lemma \arabic{lemma}. }
}
\def\endlemmaclaim{
\par\medskip}
\newenvironment{lemma}{\lemmaclaim}{\endlemmaclaim}

\title[Treatment Effect Variation]{Randomization Inference for Treatment Effect Variation}
\author[Ding {\it et al.}]{Peng Ding, Avi Feller, and Luke Miratrix}
\address{Harvard University,
Cambridge, MA, USA.}
\email{pengding@fas.harvard.edu}

\begin{document}

\begin{abstract}
Applied researchers are increasingly interested in whether and how treatment effects vary in randomized evaluations, especially variation not explained by observed covariates. We  propose a model-free approach for testing for the presence of such unexplained variation. To use this randomization-based approach, we must address the fact that the average treatment effect, generally the object of interest in randomized experiments, actually acts as a nuisance parameter in this setting. We explore potential solutions and advocate for a method that guarantees valid tests in finite samples despite this nuisance. We also show how this method readily extends to testing for heterogeneity beyond a given model, which can be useful for assessing the sufficiency of a given scientific theory. We finally apply our method to the National Head Start Impact Study, a large-scale randomized evaluation of a Federal preschool program, finding that there is indeed significant unexplained treatment effect variation.
\end{abstract}


\section{Introduction}
Researchers and practitioners are increasingly interested in whether and how treatment effects vary in randomized evaluations. 
For example, economic theory predicts that changes in welfare policy will lead to heterogeneous responses beyond those explained by observable characteristics.
Given this, how can we use experimental data to assess whether there is in fact unexplained treatment variation~\citep{Bitler:2010ty}? 
Similarly, we might be interested in assessing the effect of scaling up a promising intervention evaluated on a limited subpopulation~\citep{OMuircheartaigh:2014wx}. If we only use observed characteristics to predict the program's effectiveness on the new population, we might wonder if we are missing critical unexplained variation, which could undermine our generalization. The goal of this paper is to build a framework to assess treatment effect variation not explained by observed covariates, also known as \textit{idiosyncratic variation}~\citep[e.g.,][]{heckmanetal1997,djebbari2008heterogeneous}.

Unfortunately, assessing such variation is difficult---to paraphrase \emph{Anna Karenina}: ``constant treatment effects are all alike; every varying treatment effect varies in its own way.'' In general, researchers investigating specific types of idiosyncratic variation must therefore rely on strong modeling assumptions to draw meaningful conclusions from the data~\citep{Cox:1984tx, heckmanetal1997, gelman2004treatment}. The key contribution of our paper is an approach that tests for the presence of unexplained treatment effect variation without requiring any such modeling assumptions.  In the simplest case, the proposed method is a test of the null hypothesis that the treatment effect is constant across all units. More generally, the approach tests whether there is significant unexplained variation beyond a specified model of treatment effect. 

Of course, all treatment effects vary in practice, especially in the social sciences, which is our area of application. The key question is whether the unexplained variation is sufficiently large to be of substantive importance. As with all omnibus-type testing procedures, rejecting this null hypothesis does not provide any indication of the source of the unexplained variation. Rather, we view this procedure as a non-parametric first step in characterizing the results of a randomized experiment. 

In the simplest no-covariate case, the goal of this approach is to test whether the treatment outcome distribution is the same as the control outcome distribution shifted by the Average Treatment Effect (ATE), a constant.
Such testing would be straightforward if this shift were known---we could simply apply standard Kolmogorov-Smirnov-type (KS) tests.
However, since the shift is not known, it is a nuisance parameter that we must estimate. 
In this case, otherwise sensible methods, such as ``plug-in'' approaches, can fail, even asymptotically~\citep[e.g.,][]{Babu:2004ve}. Incorporating covariates only compounds this problem. 

Testing features of distributions in the presence of nuisance parameters has a long history in statistics and econometrics, where in the latter it is known as the \textit{Durbin Problem}~\citep{durbin1973distribution}. 
Several papers tackle this issue in the context of comparing treatment and control outcome distributions, appealing to various asymptotic justifications to bypass the nuisance parameter problem. These include a martingale transformation \citep{koenker2002inference} and subsampling \citep{chernozhukov2005subsampling}.

We take a different approach, exploiting the act of randomization as the ``reasoned basis for inference''~\citep{fisher1935design}. The corresponding Fisher Randomization Test (FRT) does not rely on further model assumptions, asymptotics, or regularity conditions~\citep[for a review, see][]{2002observational}. 
For the constant treatment effect case, when the ATE is assumed known, the FRT procedure yields an exact $p$-value for the sharp null hypothesis of a constant treatment effect~\citep[for one generalization, see][]{Abadie:2002tm}. When the ATE is unknown, the null hypothesis is no longer sharp. To correct for this, we  first construct a confidence interval for the ATE, repeat the FRT procedure pointwise over that interval, and then take the maximum $p$-value. As~\citet{Berger:1994vh} show, this procedure guarantees a valid test, despite the presence of the nuisance parameter. 
This process readily generalizes for testing treatment effects beyond a hypothesized model.

Our FRT-based approach has several key advantages. First, since the FRT approach is justified by the physical randomization alone, it yields valid inference in finite samples without relying on asymptotics or requiring absolutely continuous outcomes. Second, the FRT automatically accounts for complex experimental designs, such as stratified and matched-pair randomizations or even re-randomization~\citep{morgan2012rerandomization}. Third, this procedure is valid for any test statistic, though some statistics will be more powerful in certain settings. With this flexibility, researchers can easily extend the FRT approach, tailoring the specific test statistic to their particular problem of interest.

Using this framework, we assess treatment effect variation in the National Head Start Impact Study, a large-scale randomized evaluation of Head Start, a Federal preschool program~\citep{puma2010hsis}. 
After evaluating a range of null models, we find that there is substantial unexplained treatment effect variation, even when considering heterogeneity across age of student, dual-language learner status, and baseline academic skill level, suggesting that policymakers should not base key decisions on the topline results alone.

The paper proceeds as follows. Section 2 describes treatment effect variation using the potential outcomes framework as well as how variation depends on the chosen outcome scale. Section 3 gives an overview of various measures of treatment effect variation.  Section 4 outlines the FRT method we propose, and Section 5 generalizes this approach to incorporate covariates. Sections 6 through 8 provide some simulation studies, apply this approach to Head Start, and discuss next steps. The on-line supplementary material contains all proofs as well as additional details.

\section{Defining Treatment Effect Variation}

Following the causal inference literature, we describe our approach using the potential outcomes framework~\citep{neyman90, Rubin:1974wx}.
We focus on the case of a randomized experiment with a binary treatment, $Z_i$, and continuous outcome, $Y_i$. 
Let $N$ be the number of subjects in the study, with $N_1$ of them randomly assigned to treatment and $N_0$ of them assigned to control. 
As usual, we invoke the Stable Unit Treatment Value Assumption, which states that there is only one version of the potential outcomes and that there is no interference between subjects \citep{rubin1980randomization}. 

With this setup, the potential outcomes for subject $i$ under treatment and control are $Y_i(1)$ and $Y_i(0)$. The \textit{science table} is the $N \times 2$ table containing the potential outcomes for all $N$ units \citep{rubin2005causal}.
Each individual's observed outcome is a function of the treatment assignment and the potential outcomes,
$$Y_i^{obs} = Z_iY_i(1) + (1 - Z_i)Y_i(0),$$
where the randomness comes only from the random treatment assignment. 
Let $\mathbf{Z}$ and $\mathbf{Y}^{obs}$ denote the treatment assignment and observed outcome vectors, respectively.
We define the individual treatment effect in the usual way as $\tau_i = Y_i(1) - Y_i(0)$, but note that other contrasts are also possible.
Finally, we define the finite sample average treatment effect as: 
$$
\tau = \frac{1}{N} \sum_{i=1}^N Y_i(1) - Y_i(0).
$$
This is a statement about the $N$ units we observe. In other words, we condition on the sample at hand.

The treatment effect is constant if $\tau_i = \tau$ for all $i=1,\ldots,N$. Otherwise, we say that the treatment effect varies across experimental units. In the language of hypothesis testing, we can define the constant treatment effect null as:
\begin{equation}
H_0^{C} : Y_i(1) -  Y_i(0) = \tau \qquad \forall i \qquad \text{ for  some } \tau.  \label{eq:null_sharp}
\end{equation}
If $\tau$ were known to be $\tau = \tau_0$, this hypothesis becomes sharp.

\subsection{Constant shift}
We can not, however, directly observe any individual-level treatment effects, $\tau_i$, since we only ever observe one potential outcome for each unit.
Instead, we observe the marginal distributions of the treatment and control groups.
Because of this, much of the literature \citep[see, e.g.,][]{Cox:1984tx} defines a ``constant treatment effect'' as a statement that the marginal CDFs of the potential outcomes of the experimental  and control unit distributions $F_0(y)$ and $F_1(y)$  are a constant shift apart:
\begin{equation}
 H_0: {F}_1(y) = {F}_0(y - \tau) \qquad \mbox{ for some } \tau. \label{eq:null_cdf} 
\end{equation} 
Rejecting $H_0$ implies rejecting the more restrictive null $H_0^{C}$ that $\tau_i = \tau$ for all $i$, but rejecting $H_0^{C}$ does not necessarily imply rejecting $H_0$.
That said, it is difficult to imagine the practical situation in which there is a substantial, varying treatment effect that nonetheless yields parallel CDFs. 
Even more interestingly, the two nulls appear to be indistinguishable given observed data.
Therefore, while not formally correct, we generally view tests for $H_0^C$ as tests for $H_0$.
Simulation studies, not shown, suggest that this practice generally leads to valid, if somewhat conservative, tests.
Understanding this relationship is an important area of future work, and is closely related to the interplay between Neyman- and Fisher-style tests.  See, for example, \citet{ding:neymanfisher2014}.

\subsection{Treatment effect variation and scaling}
Whether a given treatment effect is constant critically depends on the scale of the outcomes. For example, a job training program that has a constant effect in earnings does not have a constant effect in log-earnings. This scaling issue is a particularly salient issue if the outcome is, say, test scores in an educational context where scale is not necessarily well defined.

\citet{Cox:1984tx} demonstrates the importance of scaling in a special case first explored by G.E.H. Reuter: if the marginal CDFs of $Y(1)$ and $Y(0)$ do not cross, there exists a monotone transformation such that the distributions of the transformed treatment and control outcomes are a constant shift apart. Unfortunately, G.E.H. Reuter has since passed away and his proof is lost to the literature; we provide a proof of this theorem in the supplementary material. 
\begin{thm}{}
Assume $F_1(\cdot)$ and $F_0(\cdot)$ are both continuous and strictly increasing CDFs of the marginal distributions of $Y(1)$ and $Y(0)$, respectively, with strict stochastic dominance $F_1(y) < F_0(y)$ for all $y$ on $[F_0^{-1}(0), F_1^{-1}(1)]$. There exists an increasing monotone transformation $g$ such that the CDFs of $g\{ Y(1)\}$ and $g\{ Y(0) \}$ are parallel. \label{thm::scale} 
\end{thm}
\noindent While the applicability of this result is limited to non-crossing CDFs, it nonetheless emphasizes the importance of scale and of understanding the problem at hand. In general, whether a given transformation is substantively reasonable depends on the context: a cube-root transformation might be very sensible if the outcome is in $cm^3$, but not if the outcome is in dollars~\citep{berrington2007interpretation}.

\section{Measures of Treatment Effect Variation}
\label{sec:test_statistic}

There are many approaches to measuring treatment effect variation, dating back to early work on non-additivity in randomized experiments~\citep[see][]{berrington2007interpretation}. We briefly highlight three basic measures: comparing marginal variances, comparing marginal CDFs, and comparing marginal quantiles. 
The usual testing procedures with the measures discussed here typically yield reasonable inference only when particular conditions, such as Normality or asymptotic regularity, are met.
In the next section, we show how the Fisher Randomization Test can yield exact $p$-values with any of these test statistics, regardless of whether these conditions are met.

To fix notation, assume that the potential outcome for treatment $z$ is drawn from the distribution of $Y(z)$, with marginal PDF $f_z(y)$, CDF $F_z(y)$, quantile function $F_z^{-1}(q)$, mean $\mu_z$, and variance $\sigma_z^2$. 
Sample analogues are denoted with hats.

\subsection{Comparing variances} 
Following~\citet{Cox:1984tx}, we can assess treatment effect heterogeneity by examining the marginal variances of the treatment and control outcomes. In particular, if the treatment effect is constant, $Y_i(1) = Y_i(0) + \tau$ and $\var{Y_i(1)}  =  \var{Y_i(0)}$. Therefore, unequal sample variances imply treatment effect heterogeneity, although the converse is not necessarily true. This makes the variance ratio,
$$t_{var}= \frac{\widehat{\sigma}_1^2}{\widehat{\sigma}_0^2}, $$
an attractive statistic, especially if the researcher believes that the treatment plausibly induces greater variance~\citep[e.g.,][]{gelman2004treatment}.

Furthermore, if the marginal distributions of potential outcomes are Normal, then $t_{var}$ follows an $F$ distribution and the corresponding test is the Uniformly Most Powerful test of equal variance. However, as we show in the supplementary material, the $F$ test is highly sensitive to departures from Normality, even asymptotically. 
We also provide a test that uses higher-order moments, such as kurtosis, to improve inference in this case.


\subsection{Comparing CDFs}
In general,  second-order moments might not capture some important features of heterogeneity, especially when $\tau_i$ varies with $Y_i(0)$. For example, a classroom intervention might have the largest effect on the lowest performing students. An alternative approach compares marginal CDFs rather than higher-order moments, suggesting the use of a KS-like test to compare the treatment and control groups. The classic KS statistic, which measures the maximum point-wise distance between two curves, is
$t_{KS} = \max_{y} |\widehat{F}_1(y) - \widehat{F}_0(y)| $.
This test, however, could reject if the treatment effect is constant but non-zero, since it is an omnibus test for any difference in distribution. 

To focus on heterogeneous treatment effects, we want to shift one of the CDFs by the ATE, and then compare the resulting distributions. In particular, if $\tau$ were known, we could calculate:
\[ t_{KS}(\tau) = \max_y \left | \widehat{F}_0(y) - \widehat{F}_1(y + \tau) \right | . \]
Under the null, the two aligned CDFs should be the same and we can directly compare the observed test statistic to the null distribution for the classic, non-parametric distribution-free KS test. 

In practice, $\tau$ is unknown and is therefore a nuisance parameter. One natural approach is to plug in the difference-in-means estimate, $\widehat{\tau} = \widehat{\mu}_1 - \widehat{\mu}_0$,  yielding the ``shifted'' KS (SKS) statistic:
$$
t_{SKS}  =  \max_y \left | \widehat{F}_0(y) - \widehat{F}_1(y +\widehat{\tau}) \right | .
$$
As we prove in the supplementary material, however, comparing this test statistic to the usual null KS distribution yields invalid $p$-values. In particular, $t_{SKS}$ converges to an asymptotic distribution that depends on the underlying distributions of the outcomes.

\subsection{Comparing quantiles}
A third approach focuses on quantiles rather than on CDFs. In this formulation,
$$F^{-1}_1(q) = F^{-1}_0(q) + \tau(q),$$
where $\tau(q)$ is the \textit{quantile process} for the treatment effect. If the effect is constant, then $\tau(q)$ is constant across $q$. Based on this,~\citet{chernozhukov2005subsampling} propose a class of test statistics based on the estimated quantile process,
$$t_{QP} = \big\Vert \widehat{\tau}(q) - \widehat{\tau} \big\Vert,$$
where $\widehat{\tau}(q)$ is an estimate of the treatment effect at the $q$th quantile, and $\Vert \cdot \Vert$ is some norm, such as \textit{sup}. 

As in the CDF case, $\tau$ remains a nuisance parameter, so the Durbin problem remains.
\citet{chernozhukov2005subsampling} solve this via subsampling. Their main argument is that, under some regularity conditions, a particular form of re-centered subsampling can yield asymptotically valid tests for whether $\tau(q)$ is constant, despite dependence on $\widehat{\tau}$.~\citet{chernozhukov2005subsampling} and ~\citet{Linton:2005tb} also propose a bootstrap variant of the subsampling procedure, though the bootstrap does not have the same general, theoretical guarantees as subsampling. For other approaches for inference on quantiles, see~\citet{doksum1976plotting},~\citet{rosenbaum1999reduced}, and~\citet{koenker2002inference}.

%
%
%
%
%
%

\section{A Randomization Test for Treatment Effect Variation}
\label{sec:frt}

Our analytic approach is based on the FRT.
To perform an FRT, a researcher needs three main ingredients: a randomized treatment assignment mechanism, a sharp null hypothesis, and a test statistic, $t(\mathbf{Z}, \mathbf{Y}^{obs})$, such as those in the previous section. Under the sharp null, all missing potential outcomes can be imputed and are thus known. Given all the potential outcomes, a researcher can then enumerate the possible values of a specified test statistic under all possible randomizations. This enumeration forms the exact null distribution, called the reference distribution, of that statistic.



\subsection{FRT with known $\tau$}
First consider a sharp null hypothesis of no heterogeneity for a known $\tau$:
$$
H_0^\tau: Y_i(1) = Y_i(0) + \tau \qquad \forall i.
$$
\noindent Given this null, we can immediately impute all missing potential outcomes from the observed data.
For a unit with $Z_i = 1$, the potential outcome under treatment is $Y_i^{obs}$ and the potential outcome under control is $Y_i^{obs} - \tau$. For a unit with $Z_i=0$, the potential outcome under treatment is $Y_i^{obs} - \tau$ and the potential outcome under control is $Y_i^{obs}$.  

 The steps of the FRT are then:
\begin{enumerate}
\item Calculate the test statistic for the observed data, $t = t(\mathbf{Z}, \mathbf{Y}^{obs})$.

\item Given the observed outcomes, $Y_i^{obs}$, the treatment assignment, $Z_i$, and the sharp null, $H_0^{\tau}$, generate the corresponding science table.
	
	\item Enumerate all possible treatment assignments, $\widetilde{\mathbf{Z}}$, under the given treatment assignment mechanism.  These are all possible treatment assignments that we could have observed for a given experiment. 
	Typically, there are too many possible enumerations so we instead take a random sample from the set of all possible assignment vectors.

	\item For each possible assignment,  $\widetilde{\mathbf{Z}}$, compute: (1) the observed outcomes, $\widetilde{\mathbf{Y}}^{obs}$  given $\widetilde{\mathbf{Z}}$ and the science table; and (2) the test statistic $\widetilde{t} = t(\widetilde{\mathbf{Z}}, \widetilde{\mathbf{Y}}^{obs} ) $. The resulting distribution of $\widetilde{t}$ across all randomizations is the exact distribution of the test statistic given the units in the sample and the null hypothesis.
	
	\item Compare the observed statistic $t$ to its null distribution and obtain the $p$-value $$p(\tau) \equiv \Pr\left(~t \geq  \widetilde{t}~\right).$$
	
\end{enumerate}

\noindent This procedure yields an exact test for any test statistic assuming $\tau$ is known. 
For instance, we can use as test statistics any of the measures of treatment effect heterogeneity discussed in the previous section, such as $t_{var}, t_{SKS}$ and $t_{QP}$.


\subsection{FRT with unknown $\tau$}

When $\tau$ is unknown, the null hypothesis is no longer ``sharp'' in the sense that we can no longer impute all the missing potential outcomes. 
We provide two options.

\subsubsection{FRT plug-in method (FRT-PI)} 
One option is to impute the science table with the estimated $\widehat{\tau}$ instead of $\tau$, and run the FRT to obtain the distribution of $t$ for that table.  
Ideally, if $\widehat{\tau}$ is close to $\tau$, the resulting science tables will be close in that the exact reference distribution for the imputed science table should look similar to the true reference distribution for our sample.
If this is the case, then inference from this plug-in procedure should be close to the case where $\tau$ is known, i.e., $p(\widehat{\tau}) \approx p(\tau)$.
Nonetheless, as~\cite{Berger:1994vh} discuss, there are no general theoretical guarantees from such a procedure.
In fact, as we show in the simulation studies, this approach can lead to invalid results when $\widehat{\tau}$ is highly variable, such as for skewed distributions, though it does appear to have sensible size for approximately Normal outcomes.

Nevertheless, this approach is distinct from appealing to the asymptotic distribution of a given test statistic.
Instead, this attempts to generate a reference distribution based on the data at hand, which may make the Durbin problem far less severe. 
Even so, as we show next, we can guarantee validity with a mild extension of this approach.


\subsubsection{FRT confidence interval method (FRT-CI)} 
An alternative approach is to find the maximum $p$-value across all values of the nuisance parameter, $\tau' \in (-\infty, \infty)$:
$$p_{\sup} = \sup_{\tau'} p(\tau')$$
where $p(\tau')$ is obtained by performing an FRT under the sharp null $H_0^{\tau'}$. Although $p_{\sup}$ is conservative, it is still valid since
$\Pr(p_{\sup} \leq \alpha)\leq \Pr( p(\tau) \leq\alpha ) \leq \alpha $. This approach, however, leads to two complications in practice: (1) maximizing a quantity over the entire real line is computationally intractable; and (2) doing so can lead to a dramatic loss in statistical power. 

\cite{Berger:1994vh} propose a convenient fix to these issues---rather than maximize over the entire real line, they instead maximize over a $(1-\gamma)$-level confidence interval for $\tau$, $CI_\gamma$: 
\begin{eqnarray*}
p_\gamma = \sup_{\tau'\in CI_\gamma} p(\tau') + \gamma.\label{eq::p-ci}
\end{eqnarray*} 
Following~\citet{2002observational}, we could obtain an exact (under $H_0^C$) confidence interval, $CI_\gamma$, by inverting FRTs for a sequence of sharp null hypotheses, $Y_i(1) - Y_i(0) = \tau'$. In practice, we approximate this confidence interval based on the Neyman variance estimator~\citep{neyman90}.
The following proposition guarantees the validity of the resulting $p$-value.
 \begin{prop}{}
Given that $CI_\gamma$ is a $(1-\gamma)$-level confidence interval for $\tau$, $p_\gamma$ is a valid $p$-value, in the sense that $\Pr(p_\gamma \leq \alpha)\leq \alpha$ under the null.\label{thm::test-ci}
 \end{prop}
\noindent As \citet{Berger:1994vh} note, the behavior of the $p$-values at the tails of the nuisance parameter interval can be complex and, unsurprisingly, depends on both the specific test statistic and the value of the nuisance parameter. For example, they might climb or be driven to zero. While we cannot provide theoretical guarantees, in our experience, the $p$-values for our chosen test statistics, $t_{SKS}$, tend toward 0 or remain flat for values of $\tau'$ moderately far from $\tau$, which suggests our method does not sacrifice much in terms of power.

 \begin{figure}
	\centering
	\makebox{\includegraphics[scale = 0.70]{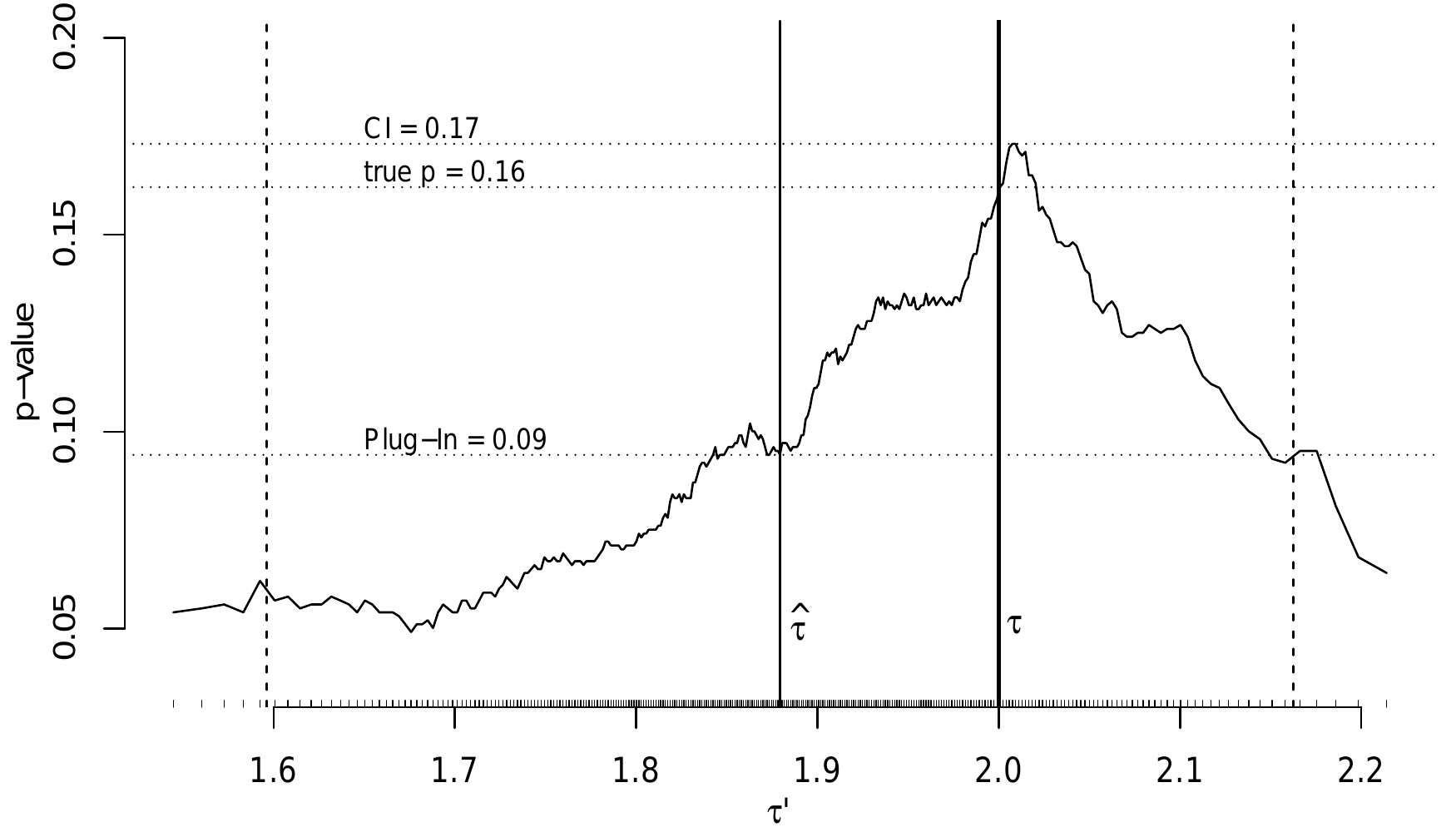}}
	\caption{$p$-values over the range of the nuisance parameter. The rug indicates the grid of sampled $\tau$.  Vertical dotted lines indicate bounds of 99.9\% confidence interval $CI_\gamma$ on $\tau$.  Horizontal lines indicate $p$-values from, bottom to top, plugging in $\widehat{\tau}$, using the known $\tau$, and maximizing the $p$-value over $CI_\gamma$.}\label{fig:ks_sim_p_values} 
 \end{figure}

To illustrate this procedure, we simulate a balanced randomized experiment with a constant treatment effect, $N = 200$, $Y_i(0) \stackrel{iid}{\sim} \text{Exponential}(1)$, and $Y_i(1) = Y_i(0) + 2$. Figure~\ref{fig:ks_sim_p_values} shows $p$-values from FRTs for a fixed data set under $H_0^{\tau'}$ for $\tau'$ in a 99.9\% confidence interval, following the procedure described above using $t_{SKS}$ as the test statistic.
If the true $\tau$ were known, we could obtain the exact $p$-value of $p = 0.16$. 
The $p$-value at the observed value of $\widehat{\tau}$ is too low, around $p = 0.09$, demonstrating why a simple plug-in approach may yield incorrect size. 
Finally, taking the maximum $p$-value over the $99.9\%$ Confidence Interval yields a $p$-value of  $p = 0.17$, only slightly larger than the true value of 0.16.
This figure, with the ``mountain'' shape, is typical for this test statistic under many data generation processes.

%
%
%
%
%
%
%

\section{Incorporating Covariates}
In practice, we typically observe a vector of individual-level pre-treatment covariates, $\mathbf{X}$, that are possibly related to the outcome.
This can help increase the power of our test and also enable exploration of variation beyond that which can be explained by $\mathbf{X}$.

\subsection{Using covariates to improve power}
To improve power we allow the chosen test statistic to account for the relationship between the covariates and outcome, such as via a linear regression of outcome on covariates and treatment with no interaction.
In the linear regression case, for example, we can generate a ``regression adjusted KS statistic.''
This statistic compares the CDFs of the residuals of a regression of $Y$ on $\mathbf{X}$ and $Z$ (with no interaction of $\mathbf{X}$ and $Z$).
Let $\widehat{e}_i = Y_i^{obs} - \widehat{Y}_i$ be the residuals of a pre-specified regression, with $\widehat{Y}_i$ being the associated predicted values.  
Then define our test statistic as
\begin{eqnarray}
\label{eq::t_RKS}
t_{RKS}  =  \max_y \left | \widehat{F}_{e1}(y) - \widehat{F}_{e0}(y) \right |
\end{eqnarray}
where $\widehat{F}_{e1}(y)$ and $\widehat{F}_{e0}(y)$ are the empirical CDFs of the residuals $\widehat{e}_i$ for the treatment and control groups, respectively.

To motivate this, consider the simple regression of $Y$ on $Z$.  
The residuals of this regression are  $\widehat{e}_i =  Y^{obs}_i - \widehat{\mu}_1$ for the treated units and $\widehat{e}_i =  Y^{obs}_i - \widehat{\mu}_0$ for the control units.
Since $\widehat{\tau} = \widehat{\mu}_1 - \widehat{\mu}_0$ and $\widehat{F}_0(y) - \widehat{F}_1(y +\widehat{\tau}) = \widehat{F}_0(y - \widehat{\mu}_0) - \widehat{F}_1(y -\widehat{\mu}_1)$, the regression-adjusted KS statistic for the simple regression of $Y$ on $Z$ is equivalent to the Shifted KS statistic: without covariates, $t_{SKS} = t_{RKS}$.


Now, by including covariates we hope to remove variation due to covariates, making variation in treatment effect more readily apparent.
As long as $\mathbf{X}$ is predictive of $Y$, regression adjustment reduces residual variation in the marginal outcomes, but cannot directly reduce variation in the treatment effect. 
In general, covariate adjustment will therefore yield more powerful test statistics. 
Importantly, since the validity of the approach is justified by randomization alone, this adjustment does not require any underlying model assumptions. 
This approach is analogous to the classical, model-assisted covariate adjustment in randomized experiments~\citep{rosenbaum2002covariance, lin.agnostic}.

We can easily repeat this approach with $t_{var}$, re-defining the test statistic via the residual variance after a regression of $Y$ on $\mathbf{X}$.
However, accounting for covariates with quantile-based statistics is more complicated, with two basic approaches in the literature.
In the conditional approach, we re-define $t_{QP}$ via the estimate of $\tau(q)$ in a quantile regression of $Y$ on both $\mathbf{X}$ and $Z$, as in~\citet{koenker2002inference}.
In the unconditional approach, we re-define $t_{QP}$ via the estimate of $\tau(q)$ in a weighted quantile regression of $Y$ on $Z$, with weights defined as a given function of $\mathbf{X}$, as in~\citet{firpo2007efficient} or~\citet{firpo2009unconditional}.

\subsection{Treatment effect variation beyond covariates}
\label{sec:beyond_covariates}
In many applications, the constant treatment effect null may be of limited scientific interest. 
Instead, we wish to investigate whether there is significant treatment effect variation beyond a particular model for the treatment. For example,~\citet{Bitler:2010ty} propose the Constant Treatment Effect Within Subgroups model, which assumes that the \textit{average} treatment effect differs across observable subgroups (e.g., by education or age group) but is otherwise constant within those subgroups.

To make this more precise, let $\mathbf{W}$ be an $n \times (k+1)$ matrix of the unit vector and $k$ pre-treatment covariates.
The unit vector corresponds to the overall average treatment effect and the covariates allow for modeled treatment effect heterogeneity. 
We then replace the null hypothesis of a constant treatment effect with the assumption that the individual-level treatment effects are a particular function of $\mathbf{W}$:
\begin{equation}
H_0^{\mathbf{W}}: \qquad Y_i(1) - Y_i(0)   =   \bm{\beta}^\top \mathbf{W}_i   \qquad \forall i \qquad \text{ for some }  \bm{\beta},  \label{eq:OLSNull}
\end{equation}
where $\bm{\beta}$ is some (unknown) vector of coefficients for $\mathbf{W}$. 
Under the null, there is some $\beta$ such that the set of $Y_i(1) - \bm{\beta}^\top \mathbf{W}_i$ yields the same CDF as the set of $Y_i(0)$.

We can easily test this hypothesis using the regression-adjusted KS statistic, $t_{RKS}$, constructed via the residuals of a regression of $Y$ on $\mathbf{W}$, $Z$, and $\mathbf{W} \times Z$.  This regression yields point estimates $\widehat{\bm{\beta}}$ with a corresponding $(k+1)$-dimensional $(100 - \gamma)\%$ confidence region.
To obtain the FRT-PI $p$-value, we simply use the science table based on $\widehat{\bm{\beta}}$.  To obtain the FRT-CI $p$-value, we must repeat the FRT procedure for each point in a potentially high-dimensional grid. We defer a detailed discussion of the estimation issues in this setting to our companion paper.

We can also extend this regression approach to account for covariates that are not assumed to interact with the treatment (i.e., those in $\mathbf{X}$ but not $\mathbf{W}$). 
Furthermore, we can allow the treatment effect model to be arbitrarily flexible, including series expansions on the covariates, such as splines or higher-order polynomials. See \cite{crump2008nonparametric} for a discussion of non-parametric estimation in this context.



\subsection{Subgroup variation}
\label{sec:subgroup_variation}
We briefly turn to the special case in which the treatment effect is assumed to vary across discrete groups. Let $Y_{ik}^{obs}$ be the observed outcome of unit $i$ in group $k$, for $i=1, \cdots, n_k$ and $k=1, \cdots, K$, with $n_{1k}$ the number of treated units in group $k$. 

For example, consider a stratified experiment, where both $n_k$ and $n_{1k}$ are fixed. 
Of course, we can always analyze a stratified experiment as if it were $K$ separate, completely randomized experiments. 
However, we can also test whether variation across strata explains the full variation in treatment effects. This corresponds to the following joint null hypothesis of stratum-specific treatment effects, $\mathcal{T} \equiv (\tau_1, \ldots, \tau_K)$:
$$
H_0^{\text{joint}}:  Y_{ik}(1) = Y_{ik}(0) + \tau_k   \qquad \forall~i, ~~\forall~k, \qquad \mbox{for some } \mathcal{T}.
$$
Under this null, the pooled CDF of the recentered-by-stratum outcomes of all the units under treatment (i.e., the residuals from outcome regressed on strata) would be the same as for control.

To test the null, we then need a measure of discrepancy between the estimates of the two CDFs as our test statistic.
Several choices are possible. First, we can use $t_{RKS}$, the regression-based test statistic above, letting $\mathbf{W}$ be a matrix of indicators for stratum membership and $\bm{\beta}$ be $\mathcal{T}$ (with no intercept). However, if the proportions of treated units differ across strata or if homoskedasticity is implausible, pooling may not be appropriate. Instead, we can post-stratify by weighting each group-by-stratum empirical CDF with weight proportional to the stratum size. The revised $\widehat{F}_{ez}$ is then
\[ \widehat{F}_{ez}(y) =  \sum_{k=1}^K \frac{n_{k}}{n} \widehat{F}_{ekz}(y) \]
where $\widehat{F}_{ekz}(y)$ is the empirical CDFs of the $Y_{ik}^{obs} - \widehat{\mu}_{kz}$ for those units in stratum $k$ with $Z_i = z$.

\begin{figure}
	\centering
	\makebox{\includegraphics[scale = 0.70]{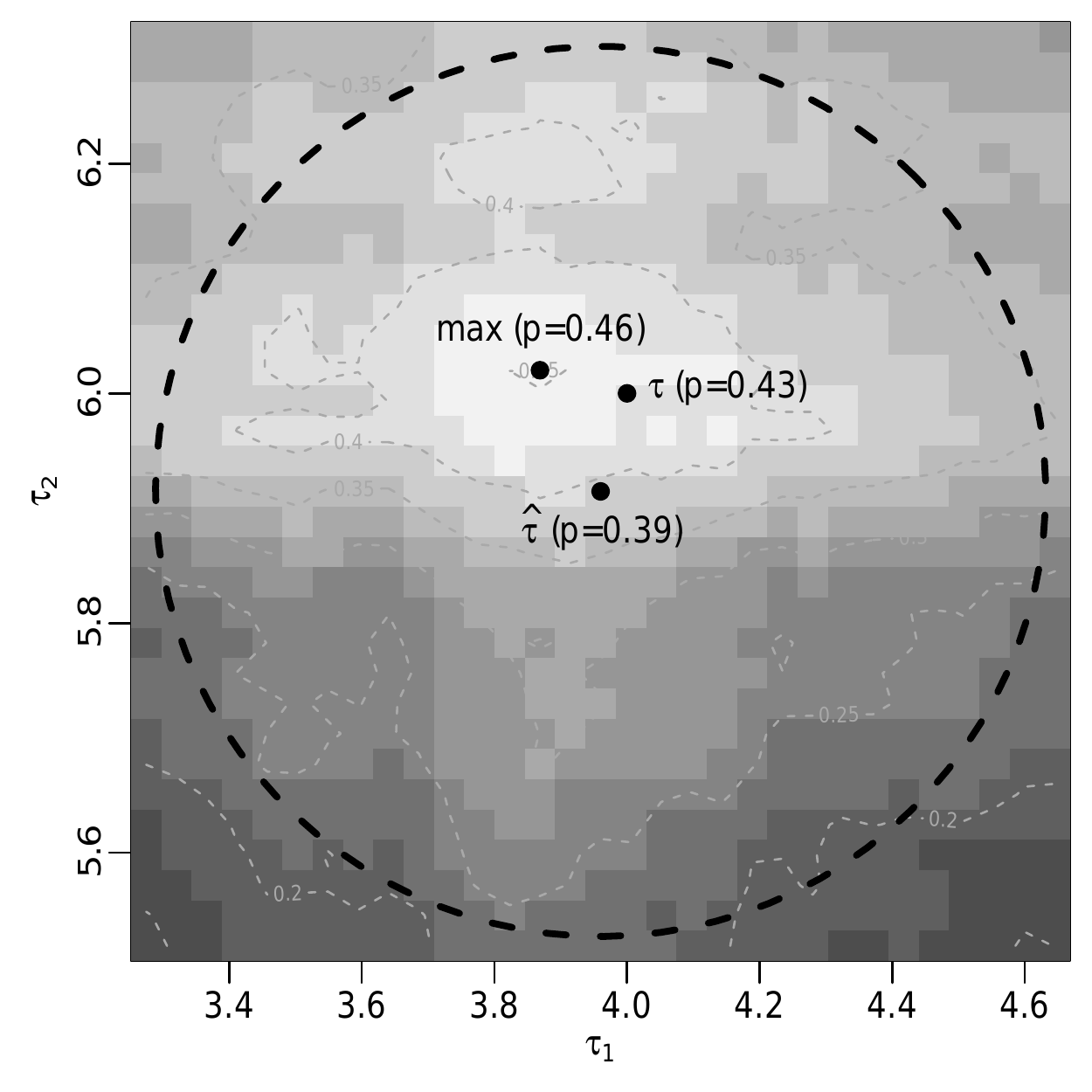}}
	\caption{$p$-values over the region of the nuisance parameters.  The $p$-values due to the maximum, the plug-in, and the oracle truth are all marked on the plot.  The dotted circle denotes the $1\!-\!\gamma$ confidence region for the nuisance parameters.}
	\label{fig:ks_sim_p_values_2d} 
 \end{figure}

Similarly, we might instead take a weighted average of individual stratum-level test statistics as 
\[ t_{WSKS} = \sum_{k=1}^K \frac{n_k}{n} t_{SKS,k} . \]
Figure~\ref{fig:ks_sim_p_values_2d} demonstrates this last approach by extending the results from Figure~\ref{fig:ks_sim_p_values} into two dimensions.  Here we have two distinct subgroups, one of 75 units and one of 375 units, and simulate a balanced randomized experiment with a treatment effect that is constant within each subgroup, but not constant overall.  The baseline distributions are exponential.
 We then test for treatment effect heterogeneity beyond these discrete subgroups.  
To do this we search over a confidence set, depicted in the figure, for a maximum $p$-value.  
We again see the ``mountain shape'' and end up with a final $p$-value of 0.46, versus $p = 0.43$ for known $\tau$ and $p = 0.39$ for the plug-in $\widehat{\tau}$. As in the one-dimensional case, the plug-in $p$-value is lower than the true $p$-value. Moreover, the maximum $p$-value is only modestly higher than the truth, as the $p$-values fall away at moderate distances from the true $\tau$. This plot is typical over several simulation settings. Finally, as expected, if we do an omnibus test for heterogeneity beyond a single average treatment effect, we reject with $p < 0.005$.
Our model of constant treatment effect within groups is thus significantly better than a single average, and we have no evidence for needing a more complex model.

The Constant Treatment Effect Within Subgroups model of~\citet{Bitler:2010ty} is equivalent to $H_0^{\text{joint}}$ except that the number of treated units in each group, $n_{k1}$, is possibly random rather than fixed.  
Here simply conditioning on the observed $n_{k1}$ for each group (i.e., only considering randomizations that maintain the $n_k$) and performing the analysis as above yields valid inference. 
This is a conditional randomization test, the analogue of post-stratification for testing rather than estimation \citep[see, e.g.,][]{Holt:1979wwa,miratrix2013adjusting}.

\section{Simulation Studies}
\label{sec:simulations}
We now turn to a series of simulation studies that confirm the validity of the FRT approach and assess power under a range of plausible scenarios. 

\subsection{Validity results}
First, we examine the different methods under the null hypothesis of a constant treatment effect. 
To assess validity, we repeat the following $5000$ times each for a given test statistic and underlying distribution:
\begin{enumerate}
\item generate a sample from the underlying distribution, assuming a constant treatment effect;
\item randomly assign treatment and obtain observed outcomes;
\item calculate our test statistic $t_{SKS}$; and finally
\item calculate a $p$-value using each of several different approaches described below.
\end{enumerate}

\noindent We assess five methods:
\begin{itemize}
	\item \textit{Naive Plug-In:} This method calculates the usual KS $p$-value, assuming that the estimated treatment effect is in fact the true treatment effect.

	\item \textit{FRT-PI and FRT-CI:} These methods are the two FRT-based approaches discussed above. For this simulation, we use a 99.9\% confidence interval for $\widehat{\tau}$ (i.e., $\gamma = 0.001$).

	\item \textit{Subsampling:} This method is the subsampling approach of~\citet{chernozhukov2005subsampling}, with the authors' recommended subsampling size of $b = 20 + n^{1/4}$. 

	\item \textit{Bootstrap:} This method is based on the bootstrap proposed by~\citet{chernozhukov2005subsampling} and~\citet{Linton:2005tb}, using the $t_{SKS}$ test statistics. To generate the bootstrap distribution, we de-mean the treatment and control groups and sample with replacement from the pooled vector of residuals, keeping the number of treatment and control units fixed.
\end{itemize}
We assess these five methods for the following distributions: standard Normal, $t_5$, standard Exponential, and Log-Normal, each with a constant treatment effect of $+1$ unit. 

Table~\ref{tab:null_coverage} shows the rejection rates for a test of size $\alpha = 0.05$ for each method and Data Generating Process (DGP). As expected, the naive plug-in approach fails dramatically, either yielding hyper-conservative or highly invalid size. The FRT-PI approach appears to work well for the symmetric Normal and $t_5$ distributions, but leads to invalid size for the skewed Exponential and Log-Normal distributions. The FRT-CI approach corrects for this, yielding exact or conservative size for all DGPs assessed here, where conservative indicates lower than nominal rejection rates. It is encouraging that, even when the FRT-CI is conservative, it is not dramatically so, suggesting we are not giving up too much power due to the maximization procedure.
Subsampling yields correct, if slightly conservative, rejection rates overall.
Finally, the bootstrap approach is invalid for the Normal, $t_5$, and Exponential distributions.

The bootstrap approach we used seemed the most promising choice.
Other alternatives to the bootstrap exist, but they seem to perform even more poorly.
For example, one seemingly obvious bootstrap is to repeatedly sample, with replacement, $N_1$ treatment cases and $N_0$ control cases from their respective original samples, calculating the resulting test statistic. 
Ideally, this would capture the variability of the entire process, giving a valid $p$-value for the actual observed test statistic. Unfortunately,  even if the null were true, the bootstrap null would generally not be in this context, and so we would end up simulating our distribution under a ``near alternative'' which gives poor size. 
We confirmed this intuition with simulations, not shown in this paper, that indeed show this approach can fail catastrophically.

\begin{table}
\caption{\label{tab:null_coverage} Size of $\alpha=0.05$ tests, in percentage points, for different methods under $H_0^C$.  Estimates are based on 5,000 replications, which implies a simulation standard error of approximately 0.3pp.} 

\centering

\begin{tabular}{r cc c cc c cc c cc}
\hline
  & \multicolumn{2}{c}{Normal} & & \multicolumn{2}{c}{$t_5$} & & \multicolumn{2}{c}{Expo} & & \multicolumn{2}{c}{Log-Normal} \\
 \cline{2-3} \cline{5-6} \cline{8-9} \cline{11-12} 
 $n=~$ &100&1,000& &100&1,000& &100&1,000& &100 &1,000\\
 \hline
\textbf{FRT-PI} &4.5&5.1& &5.4&5.2& &11.3&7.7& &15.1 &7.0\\
\textbf{FRT-CI} &1.9&3.8& &2.1&3.7& &4.1&4.9& &4.5 &5.0\vspace{5pt} \\
\textbf{Subsampling}&2.3&4.6& &1.5&1.6& &2.1&2.3& &1.1 &0.7\\
\textbf{Bootstrap} &9.3&8.8& &7.4&8.3& &6.3&6.1& &5.0 &6.5\vspace{5pt} \\
\textbf{Naive Plug-In} &0.0&0.0& &0.0&0.0& &12.3&21.4& &36.2 &44.7 \\
\hline
\end{tabular}
\end{table}

\subsection{General power simulations}
To assess the power of these methods under select alternatives we mirror a set of simulation studies conducted by both~\citet{koenker2002inference} and~\citet{chernozhukov2005subsampling}. For these simulations, we repeatedly generate data with different levels of treatment effect heterogeneity, denoted by $\sigma_\tau$, and estimate the probability that a method would reject the null of constant treatment effect (at $\alpha = 0.05$) given draws of data and random treatment assignment. 
Since the bootstrap, FRT-PI, and naive plug-in are invalid tests, we do not include them here. 

We use a binary version of the DGP from~\citet{chernozhukov2005subsampling}:
\begin{eqnarray*}
Y_i(0) = \varepsilon_i, \quad 
\tau_i = 1 + \sigma_{\tau}Y_i(0). 
\end{eqnarray*}
\noindent  with  $\varepsilon_i \sim N(0,1)$.
 This model can also be expressed as the classic additive treatment effect model under Normality,
$Y_i(1) = Y_i(0) + \tau_i$, where $Y_i(0) \sim N(0,1)$; $\tau_i \sim N\left(1, \sigma^2_{\tau} \right)$ on the margin, and $\sigma_\tau = 0$ corresponds to a constant treatment effect. Note that, as \cite{Cox:1984tx} observes, the $F$-test is the Uniformly Most Powerful Test in this setting. 

We then extend the simulations from~\citet{chernozhukov2005subsampling} by imposing Log-Normality rather than Normality. In particular, we assume a treatment effect of the following form:
\begin{eqnarray*}
\log Y_i(0) =\varepsilon_i , \quad 
\tau_i = 1 + \sigma_\tau Y_i(0).
\end{eqnarray*}

\noindent Then marginally $Y_i(1) \sim \text{Log-Normal}\left\{ \log( \sigma_\tau + 1), 1 \right\} + 1$. In either case, for $\sigma_\tau > 0$, the treatment effect increases with $Y_i(0)$, which is non-negative. \cite{rosenbaum1999reduced} calls this kind of treatment effect variation a \textit{dilated effect}.

Table~\ref{tab:cfv_sim} shows the main power results. For Normal outcomes, both the FRT-CI and subsampling methods have correct size when $\sigma_\tau = 0$. However, subsampling appears to be more powerful for $\sigma_\tau > 0$, perhaps because the the asymptotics to ``kick in'' quickly under Normality.
For Log-Normal outcomes, however, the situation is reversed, with much greater rejection rates under FRT-CI than under subsampling. 

\begin{table}
\caption{\label{tab:cfv_sim}Rejection rates for $\alpha=0.05$ tests, in percentage points, under select alternative hypotheses with different levels of treatment effect variation, $\sigma_\tau$, and DGPs.  Estimates are based on $5,000$ replications, which imply a simulation standard error of approximately 0.3pp.}
		
	\centering
	
	\begin{tabular}{r ccc c ccc}
	 \hline
	 & \multicolumn{3}{c}{\textbf{FRT-CI}}  & & \multicolumn{3}{c}{\textbf{Subsampling}}  \\
	 \cline{2-4} \cline{6-8}
	 & $\sigma_\tau = 0$ &  $\sigma_\tau = 0.2$ &  $\sigma_\tau = 0.5$ & & $\sigma_\tau = 0$ &  $\sigma_\tau = 0.2$ &  $\sigma_\tau = 0.5$ \\ 
\hline
 \multicolumn{8}{c}{ } \\
& \multicolumn{7}{c}{A. NORMAL OUTCOMES} \\

	 $N = 100$ & 2.3 & 5.5 & 23.1 & & 2.4 & 8.4 & 39.2 \\
	 $N = 400$ & 3.5 & 24.8 & 93.0 & & 4.0 & 40.1 & 98.4 \\
	 $N = 800$ & 3.5 & 52.3 & 100.0 & & 4.6 & 72.9 & 100.0 \vspace{10pt} \\
	 
& \multicolumn{7}{c}{B. LOG-NORMAL OUTCOMES} \\

	 $N = 100$ &  4.7 & 7.3 & 19.3 & & 1.2 & 2.2 & 5.9\\
	 $N = 400$ & 4.7 & 19.8 & 70.5 & & 0.6 & 3.4 & 32.9 \\
	 $N = 800$  & 4.6 & 35.1 & 94.1 & & 0.8 & 9.1 & 70.7 \\
\hline
	\end{tabular}
\end{table}

\section{Application to the Head Start Impact Study}
Initially launched in 1965, Head Start is the largest Federal preschool program today, serving around 900,000 children each year at a cost of roughly \$8 billion. The National Head Start Impact Study (HSIS) is the first major randomized evaluation of the program~\citep{puma2010hsis}. The published report found that, on average, providing children and their families with the opportunity to enroll in Head Start improved children's key cognitive and social-emotional outcomes. The report also included average treatment effect estimates for a variety of subgroups of interest, though there is only significant impact variation across a small number of the reported, pre-treatment covariates. 

After these findings were released, many researchers argued that the reported topline results masked critical variation in program impacts. For example,~\citet{Bitler:2013ve} show that the treatment is differentially effective across quantiles of the test score distribution;~\citet{Bloom2014} explore variation in program impacts across select subgroups and across the 351 Head Start centers in the study; and~\citet{Feller2014} investigate differential effects based on the setting of care each child would have received in the alternative treatment condition.

All of these approaches, however, estimate treatment effect variation by relying on a specific set of models, such as quantile or hierarchical regression. Given the breadth of research in this area, a natural question is whether the topline and subgroup average treatment effects for HSIS are indeed sufficient summaries of the program's effect. We investigate this question by focusing on the Peabody Picture Vocabulary Test (PPVT), a widely used measure of cognitive ability in early childhood. We also utilize a rich set of pre-treatment covariates, including pre-test score, child's age, child's race, mother's education level, and mother's marital status. In addition, we follow the experimental design and ensure that the randomizations used in the FRT procedure are stratified by Head Start center. For the sake of exposition, we restrict our analysis to a complete-case subset of HSIS, with $N_1 = $ 2,238 in the treatment group and $N_0 = $ 1,348 in the control group. Note that this restriction could lead to a range of inferential issues which we do not explore here; see~\citet{Feller2014} for a detailed discussion. 

As shown in Table~\ref{tbl:hsis}, we apply the FRT procedure to a set of increasingly flexible null hypotheses. The least flexible models, Model 1 and 2, assesses the null hypothesis of constant treatment effect across all units without and with covariate adjustment, using the $t_{SKS}$ statistic and the $t_{RKS}$ statistic respectively. Model 3 adjusts for pre-treatment covariates and allows the treatment effect to vary by child's age (three vs. four years old). The most flexible model, Model 4, allows the treatment effect to vary by child's age, child's Dual-Language Learner (DLL) status, and an indicator for whether the child was in the bottom quartile on an assessment of pre-academic skills prior to the study.
The resulting $p$-values are roughly $p = 0.03$ for the model without covariates and $p < 0.01$ across all three models that adjust for covariates, clearly demonstrating significant unexplained variation regardless of the exact specification. This provides evidence that there is indeed substantial treatment effect variation beyond that explained by these subgroups.

\begin{table}
		\caption{\label{tbl:hsis} FRT $p$-values for the Head Start Impact Study, based on 2,000 repetitions. Models (1) and (2) correspond to a null hypothesis of constant treatment effect. Models (3) and (4) allow the treatment effect to vary across given covariates.}
		
	\centering
	
	\begin{tabular}{rc cccc}
\hline
	&& (1) & (2) & (3) & (4) \\
\cline{3-6}
\textbf{$p$-value:} && \textbf{0.033} & \textbf{0.005} & \textbf{0.005} & \textbf{0.003} \\[0.05em]
\multirow{3}{*}{\textit{Treatment effect varies by:}} && \multirow{3}{*}{---} &  \multirow{3}{*}{---} &  \multirow{3}{*}{age} & age \\
&& & & & DLL status \\
&& & & & acad. skills \\[0.05em]
\textit{Control for covariates:} && --- & $\checkmark$ & $\checkmark$ & 	$\checkmark$  \\[0.5em]
\hline
	\end{tabular}
\end{table}

\section{Discussion} 
Researchers are increasingly interested in assessing treatment effect heterogeneity. We propose a framework to unify and generalize some existing statistical procedures for inference about such variation, using randomization as the ``reasoned basis for inference'' for the testing procedure. As a result, the method does not rely on any further model assumptions, asymptotics, or regularity conditions. We use simulation studies to confirm that this approach yields valid results in finite samples and that its power is competitive with some existing approaches, especially subsampling. Finally, we apply this method to the National Head Start Impact Study, a large-scale randomized evaluation, and find that there is indeed significant unexplained treatment variation.  

Other randomization-based approaches to heterogeneity also exist.
These methods typically specify a model for heterogeneity and test based on that model.
For example \citet{Rosenbaum:2011bu} provides randomization tests for rare but substantial effects.~\citet{rosenbaum1999reduced} proposes a randomization-based procedure for non-negative and non-decreasing quantile treatment effects under the assumption of rank preservation. See section 2.4.4 of \citet{Rosenbaum:2010book} for discussion of testing general null hypotheses of non-zero treatment effects. By contrast, we attempt to test for heterogeneity in an unstructured way, though the choice of test statistic is motivated by the problem at hand. As additional assumptions on the structure of the heterogeneity will increase statistical power, using these approaches may be more appropriate than our omnibus method when such assumptions are met.

There is one important complication that we do not directly address here: the case of discrete outcomes. Even though the FRT procedure still yields valid inference in this setting, the constant treatment effect hypothesis may no longer be of scientific interest. This is a fundamental issue and is not specific to any particular testing procedure. 
For example, consider a semi-continuous outcome distribution, with a large point mass at zero. For example, in the Connecticut Jobs First evaluation, roughly half the sample has no earnings~\citep{Bitler:2006vl}.  Here, the constant effect null hypothesis implies that welfare reform has the same dollar impact regardless of whether the individual starts with zero earnings, which is nonsensical. In future work, we hope to explore different approaches for this setting, including latent variable formulations.

In the end, our approach offers a flexible framework for assessing treatment effect variation in randomized experiments, allowing researchers to incorporate a broad range of test statistics and accommodate complex experimental designs. Most of all, our goal is to give applied researchers a set of tools so that inference about treatment effect variation can becomes standard step in the analysis of randomized experiments. 
Next steps are to explore the role of covariates in treatment effect variation and, in particular, the interplay between systematic and idiosyncratic treatment effect variation.

\section*{Acknowledgements}
The authors would like to thank Alberto Abadie, Marianne Bitler, Paul Rosenbaum, Don Rubin, Tyler VanderWeele, and participants at the Atlantic Causal Inference Conference, the Joint Statistical Meetings, and the Harvard--MIT Econometrics Workshop for helpful comments. We especially want to thank Sir David Cox for his insights and for bringing G. E. H. Reuter's lost proof to our attention. We also thank the editor and two anonymous reviewers for their very helpful feedback. The research reported here was partially funded under cooperative agreement \#90YR0049/02 with the Agency for Children and Families (ACF) of the U.S. Department of Health and Human Services. The opinions expressed are those of the authors and do not represent these institutions.


\bibliographystyle{chicago}
\begin{footnotesize}
\begin{singlespacing}
\bibliography{references2}
\end{singlespacing}
\end{footnotesize}

\bigskip 
\bigskip

\begin{flushleft}
\sffamily\bfseries\Large Supplementary Materials for\\
``Randomization Inference for Treatment Effect Variation''
\end{flushleft}

\renewcommand {\theequation} {A.\arabic{equation}}
\renewcommand {\thelemma} {A.\arabic{lemma}}
\renewcommand {\thesection} {A.\arabic{section}}

\setcounter{section}{0}
\setcounter{equation}{0}

\bigskip 
\bigskip

The supplementary materials contain four sections and a few additional notes. 
Section \ref{sec::reuter-theorem} discusses the importance of scaling in treatment effect variation, and provides a ``lost proof'' of Reuter's Theorem \citep{Cox:1984tx}.
Section \ref{sec::var-ratio} discusses the asymptotic distribution of the variance ratio statistic for non-Normal outcomes and provides a distribution-free test.
Section \ref{sec:cdf_tests} explains why the simple plug-in approach using the KS test fails, even asymptotically.
Section \ref{sec::prop1} gives a proof for Proposition 1 in the main text.

\section{Reuter's Theorem}
\label{sec::reuter-theorem}

%

As mentioned in Section 2 of the main text, whether a given treatment effect is constant depends on the scale of the outcomes.
In particular, if the marginal CDFs of $Y(1)$ and $Y(0)$ do not cross, there exists a monotone transformation such that distributions of the transformed treatment and control outcomes are a constant shift apart as defined by $H_0$ (i.e., we can make the CDFs of the treatment outcomes and control outcomes parallel).
This was first observed by \cite{Cox:1984tx}, citing a theorem due to G. E. H. Reuter.  
Unfortunately, Reuter has since passed away and his proof has been lost to the literature. We therefore provide a proof of this theorem here.
For convenience we restate the theorem:
\begin{thm}{}
Assume $F_1(\cdot)$ and $F_0(\cdot)$ are both continuous and strictly increasing CDFs of the marginal distributions of $Y(1)$ and $Y(0)$, respectively, with strict stochastic dominance $F_1(y) < F_0(y)$ for all $y$ on $[F_0^{-1}(0), F_1^{-1}(1)]$. There exists an increasing monotone transformation $g$ such that the CDFs of $g\{ Y(1)\}$ and $g\{ Y(0) \}$ are parallel. \label{thm::scale}
\end{thm}
First, note that stochastic dominance occurs if and only if the CDFs do not cross, hence the statement of the theorem above.
Next, assuming no ties, the finite sample analogue, or the analog conditioning on the realized sample, is immediate.
To prove the theorem we first need a few lemmas.

\begin{lemma}{}
Assume that $v(u)$ is a strictly increasing and continuous function defined on $[0, 1]$, which satisfies $v(u) < u$ for all $u$. Then there exists a monotone increasing transformation $h(\cdot)$ such that $h(u) - h\{  v(u) \} $ is a constant for all $u \in [0,1]$.\label{lemma::transformation}
\end{lemma}

{\sc Proof of Lemma \ref{lemma::transformation}.}
Define $v^{(1)}(u) = v(u)$ and $v^{(n)}(u) = v  \{  v^{(n-1)}(u) \}$ for $n\geq 2$; furthermore, define
$v^{(-1)}(x) = v^{-1}(x)$ and $v^{(-n)}(x) = v^{-1}\{  v^{(-(n-1))}(x) \}$ for $n\geq 2.$  

For the trivial case with $v(1) < 0$, we can easily rescale the range of $v(x)$ to make it parallel to $x$ because the range is entirely below $0$.  To do so, define $h(\cdot)$ as follows:
\[
   h(x) = \left\{
     \begin{array}{ll}
       x & \text{ if } x > v(1),  \\
       v^{-1}(x) - \{1-v(1)\} &  \text{ if } x \leq v(1).
     \end{array}
   \right.
\]
Therefore, we have
$ h(u)  - h\{ v(u)\} = u - v^{-1} \{  v(u) \} +\{1-v(1)\} = 1-v(1) $ for all $u.$ 
Note that $h(u) = u$ for all our $x \in [0,1]$ so $h$ only impacts the $h\{  v(u) \}$ term in the difference.

For the case with $v(1) \geq 0$,
we first need to show that there exist an $M$, such that $v^{(n)}(1) \geq 0$ for all $n<M$ and $v^{(M)}(1) < 0$.
We use a proof by contradiction, and assume $v^{(n)}(1) \geq 0$ for all $n.$  A quick induction shows that, since $v(1) < 1$ and 
$v^{(n+1)}(u) = v\{  v^{(n)}(u) \} < v^{(n)}(u)$, the sequence $ v^{(n)}(1) $ is strictly decreasing.
Furthermore, because the sequence is bounded below by $0$, it must have a limit, $u_0$, on $[0,1]$. 
Taking $n\rightarrow \infty$ on both sides of $v^{(n)}(u) = v  \{  v^{(n-1)}(u) \}$, we have $u_0 = v(u_0) < u_0$, which is impossible. Therefore, such an $M$ indeed exists.  We also have $v(0) \leq v^{(M)}(1)$ from taking $v$ on both sides of $0 \leq v^{(M-1)}(1)$, and we can therefore partition the real line $\mathcal{R}$ as follows:
$$
\mathcal{R} = \left(-\infty , v^{(M)}(1) \right] \cup \left( v^{(M)}(1),  v^{(M-1)}(1) \right] \cup \cdots \cup \left(v^{(2)}(1), v(1) \right] \cup \left(v(1), \infty \right).   
$$
We will define $h(\cdot)$ on each piece of the partition above, from the right to the left.

First define $h(x) = x$ within $(v(1), \infty ) $, and
$h(x) = v^{-1}(x) - \{1-v(1)\} $ within $(v^{(2)}(1), v(1)]$.
This way of construction guarantees that, for any $u\in (v(1), 1]$,
\[ h(u) - h\{  v(u) \} = u - v^{-1} \{  v(u) \} +\{1-v(1)\} = 1-v(1) . \]
Next define $h(x) = v^{(-2)}(x) - 2\{ 1- v(1)\} $ within $(v^{(3)}(1), v^{(2)}(1)]$ giving, for all $u \in (v^{(2)}(1), v(1)]$,
\[ h(u)   - h\{  v(u) \} = [  v^{-1}(u) - \{1-v(1)\} ]  - [   v^{-2}\{ v(u) \} - 2\{1-v(1)\}   ] = 1-v(1) . \]
Analogously, we can sequentially define $h(x) = v^{(-n)}(x) - n\{ 1-v(1)\}$ within all intervals $( v^{(n+1)}(1), v^{(n)}(1)  ]$ for $n < M$ which guarantees that 
\[ 
h(u)   - h\{  v(u) \} =  [  v^{(-(n-1))}(u) - (n-1)\{ 1-v(1)\} ]  -[  v^{(-(n-1))}(u) - n\{ 1-v(1)\}  ] = 1-v(1) 
\]
for all $u\in ( v^{(n)}(1), v^{(n-1)}(1)  ]$.

We finally define $h(x) = v^{(-M)} (x) - M\{1-v(1)\}$ for $x\in  ( -\infty , v^{(M)}(1) ]$, which guarantees that 
\[ 
h(u) - h\{  v(u) \}  =  [  v^{(-(M-1))}(u) - (M-1)\{ 1-v(1)\} ]  - [  v^{(-M)}(v(u)) - M\{ 1-v(1)\} ]  = 1-v(1) 
\]
for all $u \in [ 0,  v^{(M-1)}(1) ]$. 

This constructed $h(\cdot)$ satisfies $ h(u) - h\{  v(u) \} = 1-v(1)$ for all $u\in \mathcal{R}$.  $\Box$

\medskip

\begin{lemma}{}
Take random variables $A$ and $B$ with invertible CDFs $F_A(\cdot)$ and $F_B(\cdot)$.  Then 
\[ F_A(y) = F_B(y - \tau) \qquad \forall y \in [F_A^{-1}(0),F_A^{-1}(1)]  \]
is equivalent to
\[ F_A^{-1}(p) - F_B^{-1}(p) = \tau \qquad \forall p \in [0,1] . \]\label{lemma::twoforms}
\end{lemma}
\bigskip
{\sc Proof of Lemma \ref{lemma::twoforms}.}
The conclusion follows if we couple $A$ and $B$ through a common Uniform$(0,1)$ random variable $U$: $A = F_A^{-1}(U)$ and $B = F_B^{-1}(U)$.
$\Box$

\bigskip
{\sc Proof of Theorem \ref{thm::scale}.}
Define
\[
   f(x) = \left\{
     \begin{array}{ll}
       F_1(x) & \text{ if } F_1^{-1}(0) \leq x \leq F_1^{-1}(1)\\
       x - F_1^{-1}(0) & \text{ if } x < F_1^{-1}(0),
     \end{array}
   \right.
\]
which is strictly monotonic and continuous.  Furthermore, $f < F_0$ everywhere.
If $F_1^{-1}(0)$  is $\infty$ then drop the second piece from the definition.  Now, define two new random variables $A= f\{ Y(1) \}$ and $B = f\{ Y(0) \}$. Then $A\sim$ Uniform$(0,1)$, and the quantile function of $B$, $F^{-1}_B(p) = f\{  F_0^{-1}(p) \}$ is strictly increasing.
Now, because for all $p \in [0,1]$
\[ v(p) \equiv F_{B}^{-1}(p) = f\{  F_0^{-1}(p) \} < F_0\{  F_0^{-1}(p) \} = p, \]
we can Lemma 1 to obtain $h(\cdot)$ such that
\[  h( p ) - h\{  F_{B}^{-1}(p) \} = \tau \qquad \forall p \in [0,1] \]
for some $\tau$.
We can instead can show that $g( y ) \equiv h \circ f( y )$ is our transform. Define $A' = g(A)$ and $B' = g(B)$.  Then $F_{A'}^{-1} = g\left(  F_1^{-1} \right)= h$,$F_{B'}^{-1} = g\left( F_0^{-1} \right) = h\{  f( F_0^{-1} ) \} = h\left( F_{B}^{-1} \right)$, and 
\[  F_{A'}^{-1}( p ) -  F_{B'}^{-1}( p ) = \tau \qquad \forall p \in [0,1] \]
which implies that the CDFs of $A'$ and $B'$ are parallel according to Lemma 2. $\Box$

\section{Testing for Idiosyncratic Variation with Variance Ratios}
\label{sec::var-ratio}




%

If $Y_i(1)$ and $Y_i(0)$ are Normally distributed as  $\mathcal{N}(\mu_1, \sigma^2_1)$ and $\mathcal{N}(\mu_0, \sigma^2_0)$, then under the null $\sigma_1^2 = \sigma_0^2 = \sigma^2$ and the variance ratio $t_{var}=\widehat{\sigma}_1^2 / \widehat{\sigma}_0^2$, the ratio of sample variances in the two groups, follows an $F$ distribution.
While this conclusion is not generally true for non-Normal distributions, even asymptotically, we can use higher-order moments to correct this test statistic for non-Normal outcomes.

\begin{thm}{}
Assume that $Y(z)$ has finite kurtosis $\kappa_z$.
Under the null of equal variance,
$$
t_{kvar} \equiv \frac{\log \widehat{\sigma}_1^2 - \log \widehat{\sigma}_0^2}{   \sqrt{     \left (  \widehat{\kappa}_1 - 1 \right) / N_1  + \left( \widehat{\kappa}_0 - 1 \right) / N_0       }       }  
\stackrel{d}{\longrightarrow} \mathcal{N}(0, 1), 
$$
as $N\rightarrow \infty$.
\label{thm::var-kurt}
\end{thm}


{\sc Proof.}
From the classic result \citep{lehmann1999elements}, we have
$$
\sqrt{N_t} (\widehat{\sigma}_t^2 - \sigma_t^2) \stackrel{d}{\longrightarrow} 
\mathcal{N} \left\{ 0,  \text{var}(Y_t - \mu_t)^2 = \sigma^4 (  \kappa_t - 1  )   \right\},
$$
and therefore using delta-method we have that
$$
\sqrt{N_t} (\log \widehat{\sigma}_t^2 - \log \sigma_t^2) \stackrel{d}{\longrightarrow} \mathcal{N}(0,    \kappa_t - 1     ).
$$
Under the null, $\sigma_1^2 = \sigma_0^2$, and $\log \widehat{\sigma}_1^2 - \log \widehat{\sigma}_0^2$ has the following asymptotic distribution
\begin{equation}
\frac{\log \widehat{\sigma}_1^2 - \log \widehat{\sigma}_0^2}{  \sqrt{  (  \kappa_1 - 1  ) / N_1     + (\kappa_0 - 1 ) / N_0}    } 
\stackrel{d}{\longrightarrow} \mathcal{N}(0, 1).  \label{eq:asmp_var}
\end{equation}
Due to the fact that $\widehat{\kappa}_t\rightarrow \kappa_t$ in probability and Slutsky's Theorem, we have
$$
\frac{\log \widehat{\sigma}_1^2 - \log \widehat{\sigma}_0^2}{  \sqrt{   \left(  \widehat{ \kappa}_1 - 1 \right) / N_1     + \left(  \widehat{ \kappa}_0 - 1 \right) / N_0    }   } 
\stackrel{d}{\longrightarrow} \mathcal{N}(0, 1).\Box
$$

Equation~\ref{eq:asmp_var} shows that the $t_{var}$ statistic does not necessarily have an $F$ distribution---the distribution depends on $\kappa_1$ and $\kappa_2$.  
However, by plugging estimates of these fourth moments in, we do recover an asymptotically distribution free reference distribution for our statistic.
The test statistic $t_{kvar}$ essentially replaces the Normality assumption with an assumption of finite fourth moments, which is often more plausible. 
This theorem closely follows \cite{Box:1955tw}, who use randomization theory to demonstrate that the distribution of the variance ratio depends on the kurtosis of the underlying outcomes.

\section{Why the Plug-In for the CDF Approach Fails}
\label{sec:cdf_tests}


As stated in the paper, under the null $H_0$ the CDFs of each group aligned by $\tau$ should be the same.
If $\tau$ were known, therefore, we could shift the treatment group and compare the resulting distributions via a Kolmogorov-Smirnov test:
\[ t_{KS}(\tau) = \max_y \left | \widehat{F}_0(y) - \widehat{F}_1(y + \tau) \right | . \]
Under the null, we can directly compare this observed test statistic to the null distribution from the classic, non-parametric KS test. 
This would be exact.


Unfortunately $\tau$ is unknown in practice and is therefore a nuisance parameter. 
One natural-seeming approach is to plug in the difference-in-means estimate, $\widehat{\tau} = \widehat{\mu}_1 - \widehat{\mu}_0$,  yielding the ``shifted'' KS (SKS) statistic:
$$
t_{SKS}  =  \max_y \left | \widehat{F}_0(y) - \widehat{F}_1(y +\widehat{\tau}) \right | .
$$
Comparing this test statistic to the usual null KS distribution, however, yields invalid $p$-values. 
As \cite{Babu:2004ve} note, for such tests ``the asymptotic null distribution of the test statistic may depend in a complex way on the unknown parameters.'' See also \cite[][Theorem 19.23]{van2000asymptotic}, and the relevant discussion in \cite{koenker2002inference}. We show this more formally in Theorem~\ref{thm::sks-nuisance}. 

\begin{thm}{}
Let $r = \lim_{N\rightarrow \infty}N_1/N$, $BB(\cdot)$ be a standard Brownian Bridge, and $\xi(\cdot)$ be a Gaussian Process with mean function $0$ and covariance function 
$
\text{cov}\{ \xi(x), \xi(y) \}  =  \sigma_0^2 f_0(x)f_0(y).
$ 
The limiting distribution of the SKS statistic is
$$
\sqrt{N}t_{SKS} = \sup_y\Big | \sqrt{N} \left\{     \widehat{F}_1( y+\widehat{\tau} ) - \widehat{F}_0(y)  \right\}  \Big|
 \stackrel{d}{\longrightarrow} \sup_y \Big |   \frac{1}{\sqrt{r}} SBB_1 + \frac{1}{\sqrt{1-r}} SBB_0  \Big|,
$$
 where $SBB_1(\cdot)$ and $SBB_0(\cdot)$ are independent realizations of the \textit{shifted Brownian Bridge} processes:
 $
 SBB(\cdot) \equiv BB(\cdot) + \xi(\cdot),
 $
 with covariance structure 
\[ 
\text{cov}\{ BB(x), \xi(y) \}= f_0(y) F_0(x)\{ 1\!-\! F_0(x) \}  \left[   E\{ Y(0) \mid Y(0)\leq x\} -  E\{ Y(0) \mid Y(0)  > x\}  \right] .
\]\label{thm::sks-nuisance}
\end{thm}

The correlation between $BB( \cdot )$ and $\xi(\cdot )$  is the ``complex dependence'' referred to above. Since the $SBB$ processes depend on $F_0(y)$, $t_{SKS}$ also depends on $F_0(y)$ and is therefore not distribution free. 

In other words, Theorem~\ref{thm::sks-nuisance} demonstrates that ``naively" plugging in $\widehat{\tau}$ for $\tau$ yields a test statistic with a null distribution that is not the null distribution of the classic KS statistic. Intuitively, if $\tau$ is known, the asymptotic distribution of $t_{KS}$ depends on the sum of two standard Brownian Bridges, which do not depend on $F_0(y)$. The uncertainty in $\widehat{\tau}$ changes these from standard Brownian Bridges to \textit{shifted} Brownian Bridges, which do depend on $F_0(y)$:

$$\sqrt{N}t_{SKS} 
 \stackrel{d}{\longrightarrow} \sup_y \Bigg |   \underbrace{ \left\{ \frac{1}{\sqrt{r}} BB_1(y) + \frac{1}{\sqrt{1-r}} BB_0(y)\right\}}_{\text{Standard KS Distribution}} +  \underbrace{\left\{\frac{1}{\sqrt{r}}\xi_1(y) + \frac{1}{\sqrt{1-r}}\xi_0(y) \right\}}_{\text{Additional Shift}} \Bigg|.$$
%
As we show in the simulation studies in Section 6 of the main text,  the $p$-value of this naive approach can either be inflated or deflated depending on the underlying distribution, $F_0(y)$. From our experience from simulation studies, the naive plug-in approach is conservative for symmetric distributions such as Normal and $t$ distributions, but it does not yield correct type one error for skewed distributions such as Exponential and Log-Normal distributions. Table 1 in the main text illustrates this point.

{\sc Proof of Theorem \ref{thm::sks-nuisance}.}
Under the null, the means of the outcomes under treated and control satisfy $\mu_1 = \mu_0+\tau$, the variances are the same $\sigma_1^2 = \sigma_0^2 = \sigma^2$, and corresponding PDFs then satisfy $f_1(y + \tau) = f_0(y)$.
First, we have
\begin{eqnarray*}
\sqrt{N}( \widehat{\tau}  - \tau ) 
= \frac{1}{  \sqrt{r } }   \frac{1}{ \sqrt{  N_1} } \sum_{i=1}^{N_1} \{ Y_i(1) - \mu_1\}  
- \frac{1}{  \sqrt{1 - r } }    \frac{1}{\sqrt{ N_0} } \sum_{i=N_1 + 1}^{N} \{ Y_i(0) - \mu_0\} .
\end{eqnarray*}
The difference between the shifted empirical CDFs is
\begin{eqnarray*}
\sqrt{N} \left\{     \widehat{F}_{1} (y + \widehat{\tau})  - \widehat{F}_{0}  (y)  \right\} 
&=&  \sqrt{N} \left\{     \widehat{F}_{1} (y + \tau)  - \widehat{F}_{0}  (y)  \right\} + \sqrt{N} \left\{     F_{1}(y + \widehat{\tau}) - F_1(y+ \tau)    \right\} \\
&&+ \sqrt{N } \left[    \left\{     \widehat{F}_{1} (y + \widehat{\tau}) -     F_{1}(y + \widehat{\tau})     \right\}  - \left\{    \widehat{F}_{1} (y + \tau)    -    F_1(y + \tau)      \right\}  \right] \\
&=& \sqrt{N} \left\{     \widehat{F}_{1} (y + \tau)  - \widehat{F}_{0}  (y)  \right\} + \sqrt{N} \left\{     F_{1}(y + \widehat{\tau}) - F_1(y + \tau)    \right\}  + o_P(1),
\end{eqnarray*}
where the last equality is due to the stochastic equicontinuity of the indicator function.
By definition of the empirical CDFs, we have
\begin{eqnarray*}
&&\sqrt{N} \left\{     \widehat{F}_{1} (y + \widehat{\tau})  - \widehat{F}_{0}  (y)  \right\} \\
&=& \frac{1}{  \sqrt{r } } \frac{1}{ \sqrt{N_1} } \sum_{i=1}^{N_1} \left[   I\{ Y_i(1) \leq y + \tau\}  - F_1(y + \tau) \right]  
-
\frac{1}{  \sqrt{1 - r } }  \frac{1}{ \sqrt{N_0}} \sum_{i=N_1 + 1}^{N} \left[  I\{ Y_i(0) \leq y\}  - F_0(y) \right]  \\
&&+
f_0(y)  \sqrt{N} (\widehat{\tau} - \tau) + o_P(1)  \\
&=&   \frac{1}{  \sqrt{r } }  \frac{1}{ \sqrt{N_1} } \sum_{i=1}^{N_1} \left[   I\{ Y_i(1) \leq y + \tau\}  - F_1(y + \tau) \right]  
-
\frac{1}{  \sqrt{1 - r } }  \frac{ 1}{ \sqrt{N_0} } \sum_{i=N_1 + 1}^{N} \left[   I\{ Y_i(0) \leq y\}  - F_0(y) \right]  \\
&&+
f_0(y) \left[    \frac{1}{  \sqrt{r } }   \frac{1}{ \sqrt{  N_1} } \sum_{i=1}^{N_1} \{ Y_i(1) - \mu_1 \}  - \frac{1}{  \sqrt{1 - r } }    \frac{1}{\sqrt{ N_0} } \sum_{i=N_1 + 1}^{N} \{ Y_i(0) - \mu_0\}  \right]
+ o_P(1).
\end{eqnarray*}
Since both 
$$
\frac{1}{\sqrt{N_1}} \sum_{i=1}^{N_1}  \left[   I\{ Y_i(1) \leq y + \tau\}  - F_1(y + \tau) + f_0(y)  \{ Y_i(1) - \mu_1 \}   \right] 
$$ 
and 
$$
\frac{1}{\sqrt{N_0}}  \sum_{i=N_1+1}^N   \left[   I\{ Y_i(0)  \leq y\} - F_0(y)  + f_0(y) \{ Y_i(0) - \mu_0\}  \right] 
$$ 
have the same asymptotic distribution as $SBB$ defined in Theorem \ref{thm::sks-nuisance}, we have
$$
\sqrt{N} \left\{     \widehat{F}_{1} (y + \widehat{\tau})  - \widehat{F}_{0}  (y)  \right\}
\stackrel{d}{\longrightarrow}   \frac{1}{  \sqrt{r } }  SBB_1(y) - \frac{1}{  \sqrt{1 - r } } SBB_0(y).
$$
The final conclusion follows from the symmetry of the shifted Brownian Bridge. $\Box$

\section{Proposition 1 for the FRT-CI Method}
\label{sec::prop1}

 \begin{prop}{}
Given that $CI_\gamma$ is a $(1-\gamma)$-level confidence interval for $\tau$, $p_\gamma$ is a valid $p$-value, in the sense that $\Pr(p_\gamma \leq \alpha)\leq \alpha$ under the null.\label{thm::test-ci}
 \end{prop}
 
{\sc Proof of Proposition 1.} 
The validity of the FRT-CI $p$-value is fairly immediate according to \cite{Berger:1994vh}.
First, the valid confidence interval guarantees $\Pr(\tau_0 \not\in CI_\gamma) \leq \gamma$.
Second,
given the true value of $\tau_0$, randomization test yields a valid $p$-value, which implies $\Pr\left\{  p(\tau_0)\leq \alpha - \gamma\right\} \leq \alpha - \gamma$.
Third, given the fact that $\tau_0 \in CI_\gamma$, the supremum of $p$-values over $CI_\gamma$, $\sup_{\tau\in CI_\gamma}p(\tau)$, is greater than or equal to $p(\tau_0)$. These ingredients give us
\begin{eqnarray*}
\Pr( p_\gamma \leq \alpha ) &=& \Pr(p_\gamma \leq \alpha, \tau_0 \in CI_\gamma) + \Pr(p_\gamma \leq \alpha, \tau_0 \not\in CI_\gamma)\\
&\leq & \Pr\left\{  \sup_{\tau\in CI_\gamma}p(\tau) \leq \alpha-\gamma, \tau_0 \in CI_\gamma\right\} + \Pr(\tau_0 \not\in CI_\gamma)\\
&\leq &\Pr\left\{  p(\tau_0) \leq \alpha-\gamma, \tau_0 \in CI_\gamma\right\} + \gamma\\
&\leq & \Pr\left\{  p(\tau_0) \leq \alpha-\gamma \right\} + \gamma\\
&\leq & \alpha-\gamma+\gamma= \alpha. \,\Box
\end{eqnarray*}

\end{document}